\documentclass[prb,twocolumn,showpacs,superscriptaddress,floatfix]{revtex4-2}
 \usepackage{amssymb,amsfonts,amsmath}
\usepackage{adjustbox}
\usepackage{color}
\usepackage{xr}
\usepackage{braket}
\usepackage{xcolor}

\begin{document}

\title{Emergence of a new band  and the Lifshitz transition in kagome metal ScV$_6$Sn$_6$ with charge density wave }
\author{Seoung-Hun Kang}
\affiliation{Materials Science and Technology Division, Oak Ridge National Laboratory, Oak Ridge, Tennessee 37831, USA}
\author{Haoxiang Li}
\affiliation{Materials Science and Technology Division, Oak Ridge National Laboratory, Oak Ridge, Tennessee 37831, USA}
\author{William R. Meier}
\affiliation{Materials Science and Engineering Department, University of Tennessee Knoxville, Knoxville, Tennessee 37996, USA}
\author{John W. Villanova}
\affiliation{Center for Nanophase Materials Sciences, Oak Ridge National Laboratory, Oak Ridge, TN 37831, USA}
\author{Saban Hus}
\affiliation{Center for Nanophase Materials Sciences, Oak Ridge National Laboratory, Oak Ridge, TN 37831, USA}
\author{Hoyeon Jeon}
\affiliation{Center for Nanophase Materials Sciences, Oak Ridge National Laboratory, Oak Ridge, TN 37831, USA}
\author{Hasitha W. Suriya Arachchige}
\affiliation{Department of Physics and Astronomy, University of Tennessee Knoxville, Knoxville, Tennessee 37996, USA}
\author{Qiangsheng Lu}
\affiliation{Materials Science and Technology Division, Oak Ridge National Laboratory, Oak Ridge, Tennessee 37831, USA}
\author{Zheng Gai}
\affiliation{Center for Nanophase Materials Sciences, Oak Ridge National Laboratory, Oak Ridge, TN 37831, USA}
\author{Jonathan Denlinger}
\affiliation{Advanced Light Source, Lawrence Berkeley National Laboratory, Berkeley, California 94720, USA}

\author{Rob Moore\thanks{corresponding author}}
\email{moorerg@ornl.gov}
\affiliation{Materials Science and Technology Division, Oak Ridge National Laboratory, Oak Ridge, Tennessee 37831, USA}
\author{Mina Yoon\thanks{corresponding author}}
\email{myoon@ornl.gov}
\affiliation{Materials Science and Technology Division, Oak Ridge National Laboratory, Oak Ridge, Tennessee 37831, USA}

\author{David Mandrus}
\affiliation{Materials Science and Technology Division, Oak Ridge National Laboratory, Oak Ridge, Tennessee 37831, USA}
\affiliation{Materials Science and Engineering Department, University of Tennessee Knoxville, Knoxville, Tennessee 37996, USA}
\affiliation{Department of Physics and Astronomy, University of Tennessee Knoxville, Knoxville, Tennessee 37996, USA}

\begin{abstract}
Topological kagome systems have been a topic of great interest in condensed matter physics due to their unique electronic properties. The vanadium-based kagome materials are particularly intriguing since they exhibit exotic phenomena such as charge density wave (CDW) and unconventional superconductivity. The origin of these electronic instabilities is not fully understood, and the recent discovery of a charge density wave in ScV$_6$Sn$_6$ provides a new avenue for investigation. In this work, we investigate the electronic structure of the novel kagome metal ScV$_6$Sn$_6$ using angle resolved photoemission spectroscopy (ARPES), scanning tunneling microscopy (STM), and first-principles density functional theory calculations. Our analysis reveals for the first time the temperature-dependent band changes of ScV$_6$Sn$_6$ and identifies a new band that exhibits a strong signature of  a structure with CDW below the critical temperature. Further analysis revealed that this new band is due to the surface kagome layer of the CDW structure. In addition, a Lifshitz transition is identified in the ARPES spectra that is related to the saddle point moving across the Fermi level at the critical temperature for the CDW formation.  This result shows the CDW behavior may also be related to nesting of the saddle point, similar to related materials.  However, no energy gap is observed at the Fermi level and thus the CDW is not a typical Fermi surface nesting scenario. These results provide new insights into the underlying physics of the CDW in the kagome materials and could have implications for the development of materials with new functionality.
\end{abstract}

\keywords{Kagome metal, charge density wave, Lifshitz transition}

\maketitle

\section{Introduction}
The interplay between correlated electron physics and topology offers tantalizing new functionality where emergent phenomena from many-body interactions are protected at elevated temperatures and energy scales \cite{WKrempa2014,Rau2016,Tokura2017,Keimer2017}. Theoretical investigations into the kagome lattice have found a band structure with topological Dirac bands along with high density of states from electronic flat bands and van Hove saddle points, which offer unusual electronic instabilities and an opportunity for investigating this overlap \cite{Yu2012,Kiesel2012,Wang2013,Parameswaran2013,Mazin2014,Lee2016}. Depending on the band filling and many-body interactions generating instabilities at the Fermi level, phases such as density waves and superconductivity have been predicted \cite{Kiesel2012,Wang2013,Isakov2006,Guo2009,Wen2010,Kiesel2013,Ko2009}. The band structure, band filling, and influence of electronic correlations are directly observable using ARPES \cite{Lu2012,Sobota2021}.  

Experimental observation of band renormalizations influencing magnetic Weyl fermions in Mn$_3$Sn as well as the observation of the kagome band structure, including Dirac points and flat bands, plus magnetic interactions in Fe$_3$Sn$_2$, revealed kagome physics is possible in actual materials \cite{Kuroda2017,Ye2018,Yin2018,Lin2018}. Recently, the band fillings of AV$_3$Sb$_5$ (A = K, Rb, Cs) \cite{Ortiz2019} and RMn$_6$Sn$_6$ (R = rare earth) \cite{Venturini1991,Ma2021} families of kagome materials have been discovered to be more supportive of enhanced electronic instabilities.  Flat bands and saddle points have been observed for YMn$_6$Sn$_6$ while the interplay between magnetism and topology has been shown in TbMn$_6$Sn$_6$ \cite{Li2021,Yin2020}.  The kagome electronic structure along with a charge density wave (CDW) and superconductivity have been observed in kagome CsV$_3$Sb$_5$ \cite{Ortiz2020,Neupert2022,Nakayama2021,Liu2021,Kang2022}.  

The recent discovery of a CDW in ScV$_6$Sn$_6$ offers new opportunities to understand the origins of electronic instabilities in these topological kagome systems \cite{Arachchige2022}.  In particular, the investigation of vanadium kagome layers in AV$_3$Sb$_5$ and RV$_6$Sn$_6$ (R = Y, Gd-Tm, and Lu)  has provided crucial insights into these unique properties. The stacked kagome layers in AV$_3$Sb$_5$ and two kagome sheets separated by alternating RSn$_2$ and Sn$_4$ layers per unit cell in RV$_6$Sn$_6$ have both shown the potential to exhibit exotic electronic phases.

Here we report the temperature-dependent electronic structure changes of ScV$_6$Sn$_6$, a novel kagome metal material that accommodates a CDW phase below the critical temperature (T$_c$). A Lifshitz transition is identified in the ARPES spectra that is related to the saddle point moving across the Fermi level at T$_c$. This result shows the CDW behavior may be connected to nesting of the saddle point, similar to related materials~\cite{Kang2022}. However, no energy gap is observed at the Fermi level and thus the CDW is not a typical Fermi surface nesting scenario.  In addition, our ARPES spectra, STM, and first-principles calculations identified the appearance of a new band below the CDW T$_c$  attributed to the surface kagome layer. This phenomenon has not been observed previously in kagome metal systems, and since it occurs conspicuously at the critical temperature of the CDW, it will play an essential role in understanding the CDW of ScV$_6$Sn$_6$. 

\section{Results and Discussion}
\begin{figure}[htb!]
\includegraphics[width=0.5\textwidth]{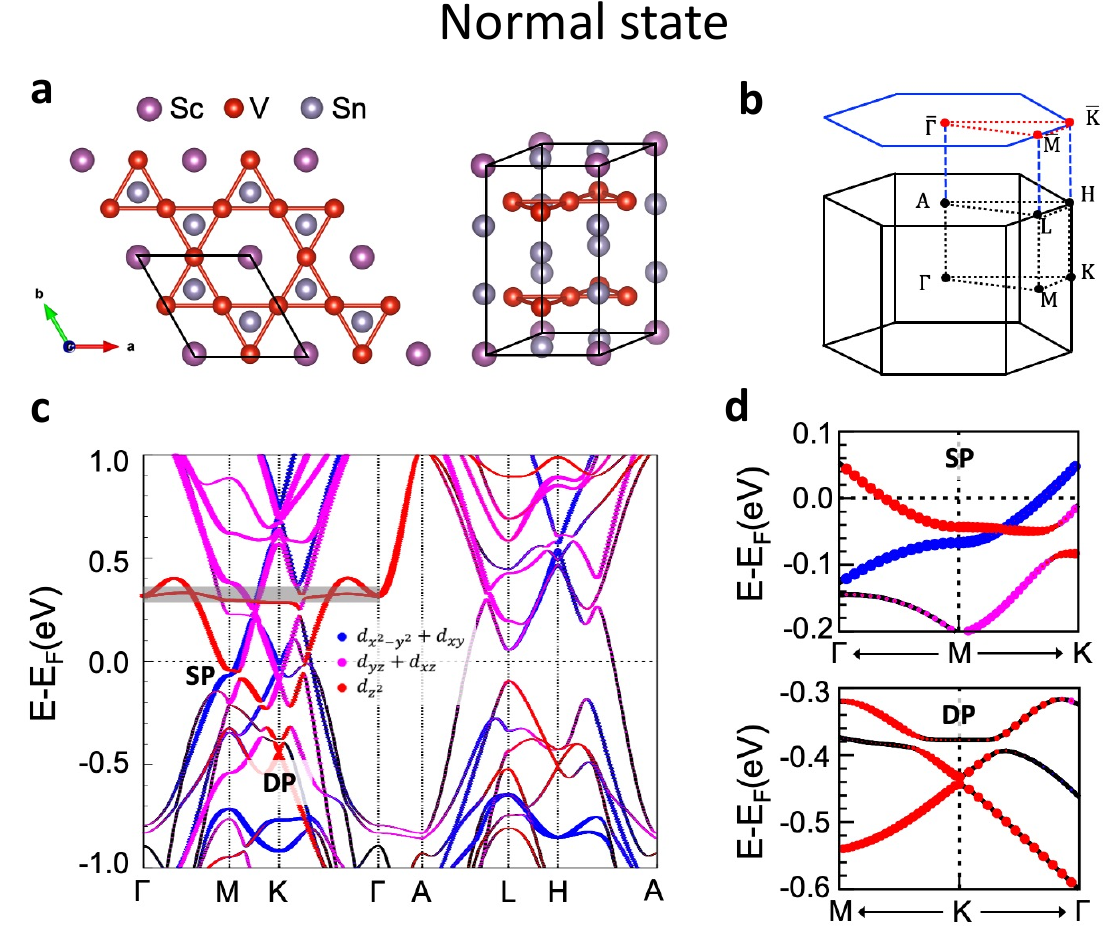}
\caption{Crystal structure and electronic band structure of kagome metal ScV$_6$Sn$_6$ in the normal state. a, Top and side views of ScV$_6$Sn$_6$ crystal structure. The kagome layers consisting of V atoms are sandwiched by ScSn and Sn atomic layers. b, Brillouin zone of ScV$_6$Sn$_6$. c, Orbital projected band structure of ScV$_6$Sn$_6$ for the d orbitals of V atoms. d, The characteristic features of the kagome lattice on ScV$_6$Sn$_6$, such as a flat band, a Dirac point (DP) at the K point, and a saddle point (SP) at the M point. The gray shaded area indicates the flat bands, the DP and the SP are marked and magnified.}
\label{fig:bulk}
\end{figure}

ScV$_6$Sn$_6$ crystallizes in the P6/mmm space group. The unit cell consists of two kagome layers composed of V atoms, which are enclosed by Sn layers and SnSc layers along the out-of-plane direction, similar to that of other AV$_6$Sn$_6$ (A=Gd, Ho, Y, Tb, etc.) \cite{Peng2021,Pokharel2021,Rosenberg2022,Pokharel2022} (Fig.~\ref{fig:bulk}a). The electronic band structures of ScV$_6$Sn$_6$, determined by first-principles density functional theory (DFT) calculations, show the characteristic features of the kagome lattice, such as a flat band due to the confinement of electrons caused by quantum interference in the kagome lattice, a Dirac point (DP) at the K point, and a saddle point (SP) at the M point from the hexagonal crystal symmetry (Fig.~\ref{fig:bulk}c,d). All bands are doubly degenerate due to the spatial inversion symmetry. Therefore the two kagome layers in the unit cell also degenerate the flat band (the gray shaded area in Fig.~\ref{fig:bulk}c). Similar features in the band structures are reported for other kagome metals~\cite{Pokharel2021, Rosenberg2022}. 

Our calculations with orbital-resolved electronic structures further confirm the one complete set of the features of the kagome lattice with the orbital composition of $d_{z^2}$. A complete set of characteristics of the kagome lattice includes a Dirac point with a small gap due to spin-orbit coupling at the K point located at 0.45~eV below Fermi level ($E_F$), a saddle point near $E_F$, and a flat band located at ~0.3~eV above $E_F$ across the entire Brillouin zone in the $k_z=0$ plane (Fig.~\ref{fig:bulk}c). 
The bands characterized mainly by the $d_{z^2}$ orbital component (red points in Fig.~\ref{fig:bulk}c) are not continuous along the M-K line due to the interaction with bands of non-orthogonal orbital compositions of $d_{xz}$ and $d_{yz}$.The importance of the orbital character of the Dirac fermions is emphasized in several other reports on kagome metals \cite{Li2021, Peng2021, Yang2022,Liu2020}.

\begin{figure}[t!]
\includegraphics[width=0.5\textwidth]{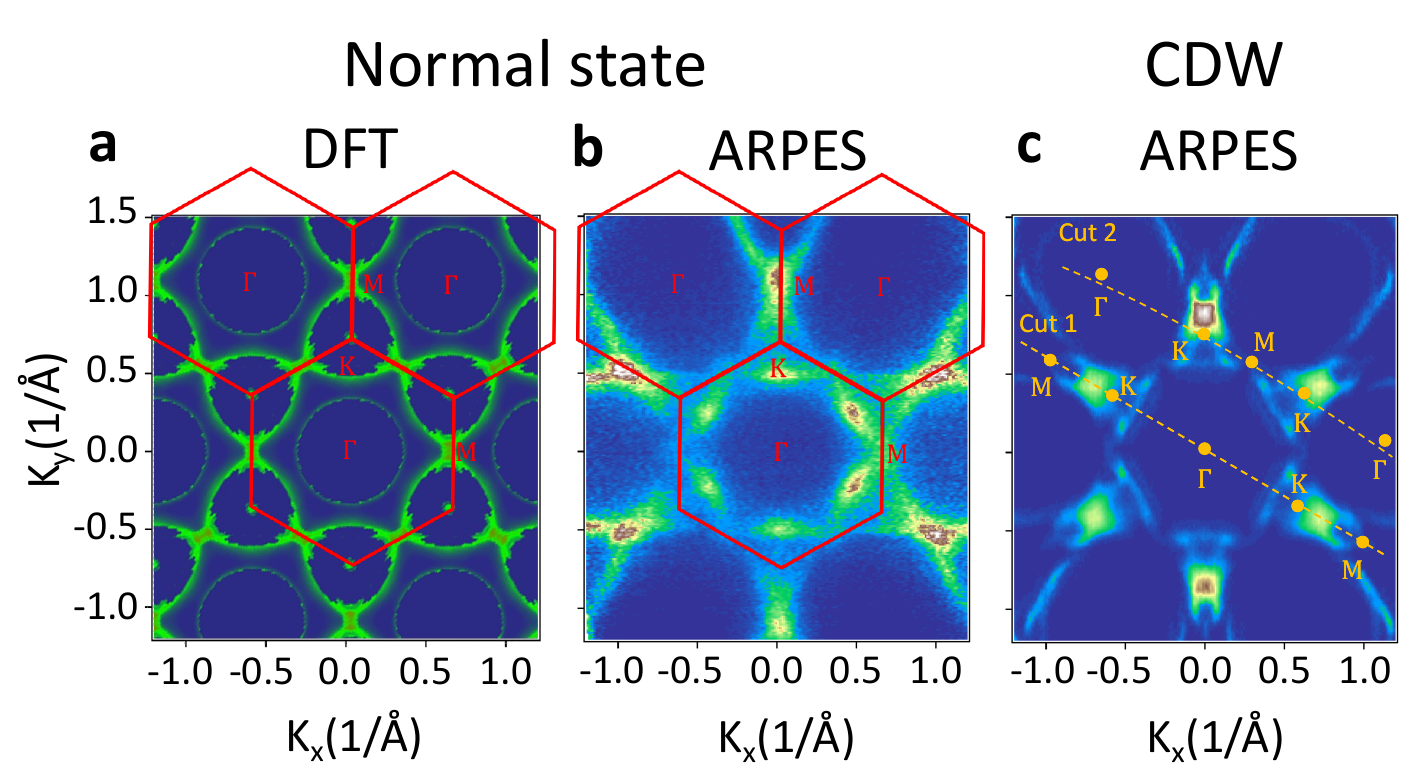}
\caption{The Fermi surfaces of ScV$_6$Sn$_6$ using DFT and ARPES. a, The theoretical Fermi surface of ScV$_6$Sn$_6$ for the normal state structure. b, The experimental Fermi surface of ScV$_6$Sn$_6$ at T = 124 K (normal state) using photon energy $h\nu$ = 98~eV with the Brillouin zone and high symmetry points highlighted. c, Experimental Fermi surface at T = 20 K (CDW state) using photon energy $h\nu$ = 98~eV with high symmetry points and high symmetry cuts shown in subsequent figures highlighted.}
\label{fig:FS}
\end{figure}

The electronic structure of ScV$_6$Sn$_6$ is investigated using DFT and ARPES with the Fermi surface in the normal state (T = 124~K) and the CDW state (T = 20~K) for the same $h\nu$ = 98~eV photon energy is shown in Fig.~\ref{fig:FS}. Interestingly, the normal state Fermi surface is similar to previous reports of the kagome termination Fermi surface for GdV$_6$Sn$_6$ while the Fermi surface in the low temperature CDW state resembles that of the Sn termination Fermi surface for GdV$_6$Sn$_6$~\cite{Peng2021}. Scanning the photon beam across the sample did not yield variations in the spectral intensity and STM images, shown in the Supplemental Information Fig.~\ref{fig:SI-STM}. This fact reveals that the ARPES data is from mixed terminations (Fig.~\ref{fig:SI-STM}a) due to the small facets from difficult-to-cleave ScV$_6$Sn$_6$ crystals. To investigate the $k_z$ dispersion, photon energy dependent scans were performed as shown in Fig.~\ref{fig:SI-kz}.  No $k_z$ dispersion is observed at the Fermi level, similar to previous measurements of YMn$_6$Sn$_6$ and RbV$_3$Sb$_5$~\cite{Li2021,Liu2021}. 

Previous investigations of ScV$_6$Sn$_6$ have revealed a CDW transition with T$_c$ = 92 K \cite{Arachchige2022}.  Neutron and X-ray studies have reported CDW structural distortions with an unusual $q_{CDW} = (\frac{1}{3},\frac{1}{3},\frac{1}{3})$ corresponding to a distinct decrease in resistivity and magnetic susceptibility on cooling through T$_c$ \cite{Arachchige2022}. 
These findings suggest changes to the Fermi surface are expected across T$_c$. While there are differences as noted above, the Fermi surfaces do not show any prominent folded electronic bands in the low temperature data due to the CDW phase. In the T = 124~K normal state Fermi surface, there is intensity at the $\bar{\text{M}}$ point as well as a point near $\bar{\text{K}}$ that lies on the $\bar{\Gamma}$-$\bar{\text{K}}$ line (Fig.~\ref{fig:FS}a,b). For the T = 20~K Fermi surface, in the CDW phase, the intensity at the $\bar{\text{M}}$ point and the intensity on the $\bar{\Gamma}$-$\bar{\text{K}}$ line is suppressed while enhanced intensity near $\bar{\text{K}}$, but on the $\bar{\text{K}}$-$\bar{\text{M}}$ line, is observed (Fig.~\ref{fig:FS}c). While this suppression of intensity could be due to electronic gaps from the CDW formation, we do not find gaps in the electronic structure at the Fermi level.  It should be noted that a similar electronic structure and similar asymmetry to the intensity on the $\bar{\text{K}}$-$\bar{\text{M}}$ line due to photoemission matrix elements has been observed for GdV$_6$Sn$_6$ where no CDW is known to exist \cite{Peng2021}. 

\begin{figure}[t!]
\includegraphics[width=0.5\textwidth]{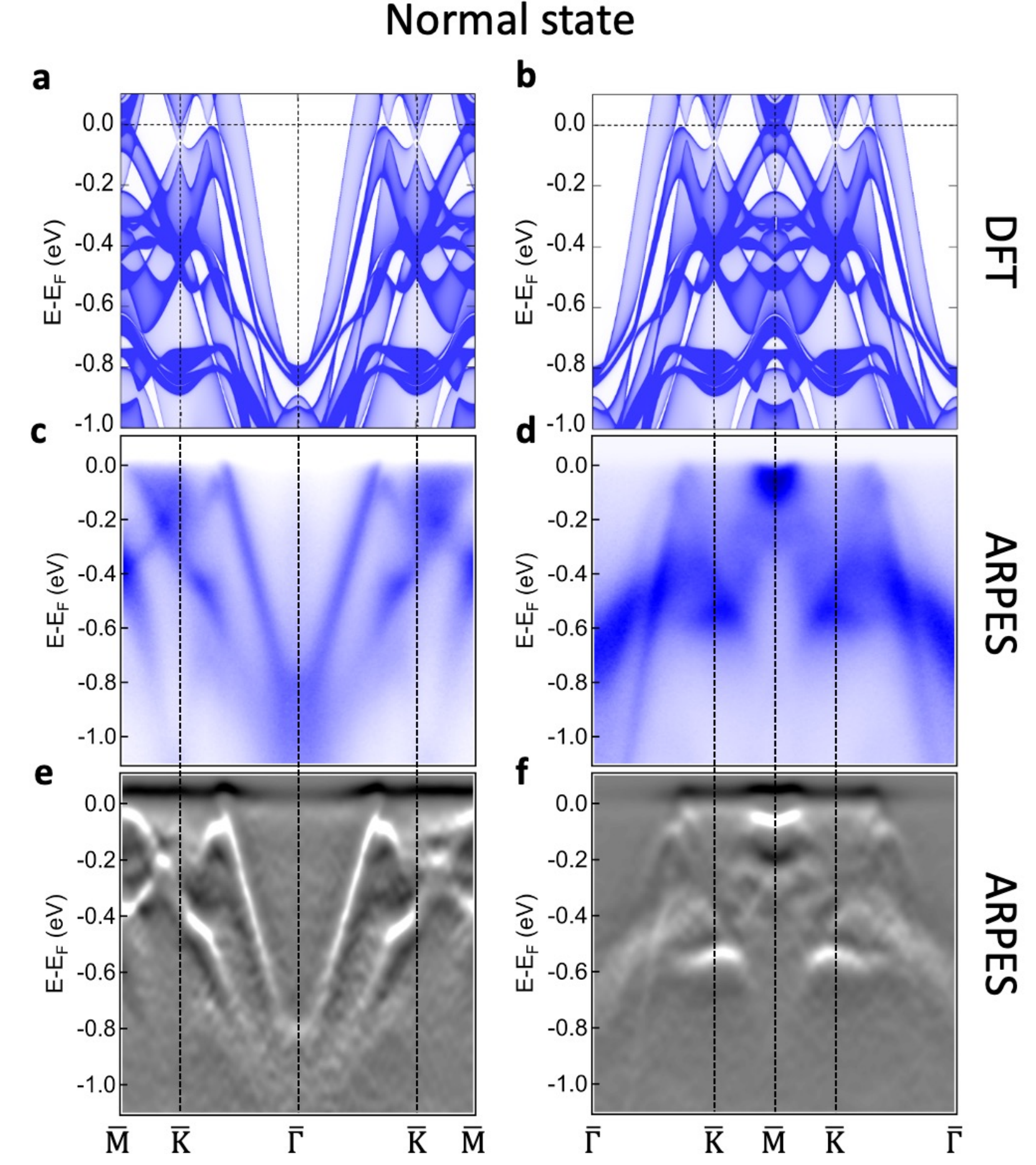}
\caption{Normal state band structures. a, DFT band structure for a semi-infinite slab along the $\bar{\text{M}}-\bar{\text{K}}-\bar{\Gamma}-\bar{\text{K}}-\bar{\text{M}}$ path and b, along the $\bar{\Gamma}-\bar{\text{K}}-\bar{\text{M}}-\bar{\text{K}}-\bar{\Gamma}$ path. c,d, ARPES band structures along the same paths and e,f, curvature analysis for the same.}
\label{fig:RTband}
\end{figure}

We further analyze the electronic structure of the high-temperature phase and show the theoretical and experimental band structures in Fig.~\ref{fig:RTband} along the two high symmetry lines defined in Fig.~\ref{fig:FS}c. Our band structures calculated for the bulk structure agree well with the band dispersion measured by ARPES. Dirac dispersion is indicated along $\bar{\text{M}}$-$\bar{\text{K}}$ at -0.2~eV below $E_F$ and the electron pocket is seen at $\bar{\text{M}}$ (Fig.~\ref{fig:RTband}a). This is seen clearly by ARPES in Fig.~\ref{fig:RTband}c,e at approximately the same energy. At $\bar{\text{K}}$ the Dirac cone at $E$-$E_F=-0.45$~eV, which is of $d_{z^2}$ character and one of the major features of the kagome band structure, is again clear in the bulk calculation (Fig.~\ref{fig:RTband}a,b), and part of the lower Dirac cone is especially bright along $\bar{\text{K}}$-$\bar{\Gamma}$ (Fig.~\ref{fig:RTband}c,e).
At $\bar{\text{K}}$ the occupied bands and electron pocket $E$-$E_F=-0.1$~eV is clear in the calculation but less prominent in the ARPES data. In Fig.~\ref{fig:RTband}c the linear Dirac dispersion of the occupied bands is faintly discernible. DFT calculations show that this Dirac dispersion and electron pocket consist of the orbital components $d_{x^2-y^2}$, $d_{xy}$, $d_{yz}$, and $d_{xz}$. 

\begin{figure*}[t!]
\includegraphics[width=1.0\textwidth]{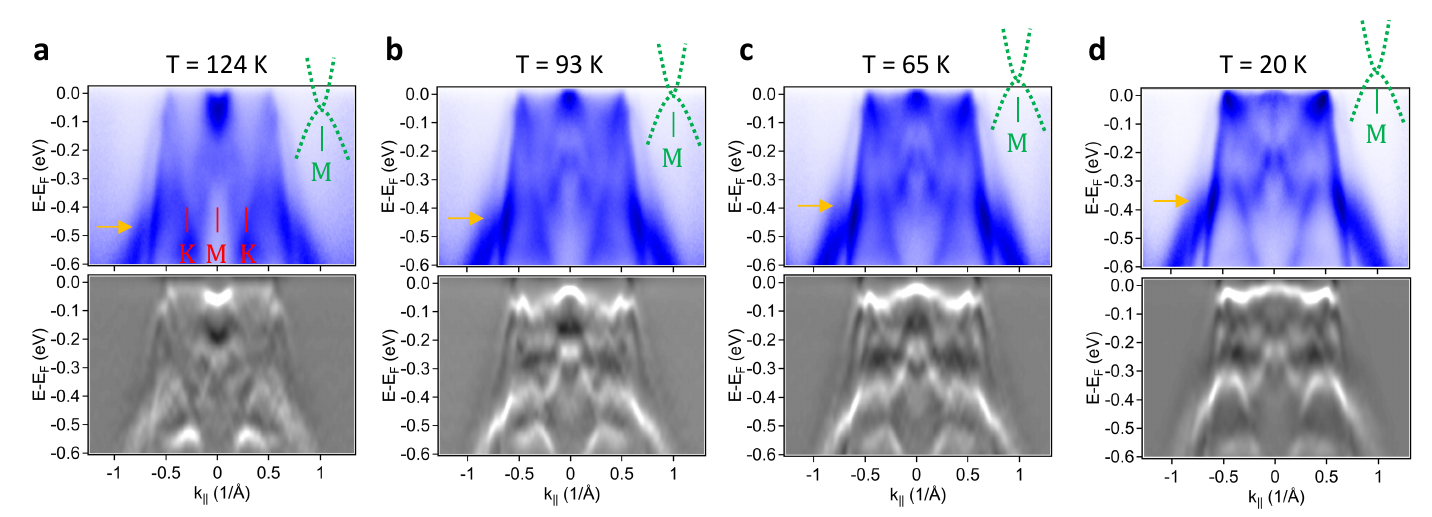}
\caption{Temperature dependence of electronic structure and and formation of the Lifshitz transition. a-d, ARPES data (top) and curvature analysis image (bottom) along Cut 2 shown in Fig.~\ref{fig:FS}c for T = 124~K, 93~K, 65~K, and 20~K, respectively.  All data taken with photon energy $h\nu$ = 82~eV. Brillouin zone high symmetry positions shown in top left panel.  Green dashed lines and $\bar{\text{M}}$  marker are guides to the eye to highlight the upward motion of the bands at the $\bar{\text{M}}$ point in the data.  Orange arrow highlights band crossing that also moves up in energy when the temperature is reduced.}
\label{fig:lifshitz}
\end{figure*}

To understand the electronic structure and its implications for the CDW, Fig.~\ref{fig:lifshitz} shows the temperature dependence of the electronic structure along Cut 2 outlined in Fig.~\ref{fig:FS}c. For these figures, centered at $\bar{\text{M}}$ along the  $\bar{\text{K}}$-$\bar{\text{M}}$ direction, a systematic upward motion in energy is observed for the band structure as the temperature is lowered across the CDW transition.  At the $\bar{\text{M}}$ point, this band shift makes the bands touching $E_F$ appear to change from an electron-like dispersion above T$_c$ to a hole-like dispersion below T$_c$, as highlighted by the green dashed curves in Fig.~\ref{fig:lifshitz}. This change in the Fermi surface contour is indicative of a Lifshitz transition due to the van Hove saddle point moving above $E_F$ as the temperature is lowered below T$_c$~\cite{Lifshitz1960}. 

For the related kagome CsV$_3$Sb$_5$ and RbV$_3$Sb$_5$ systems, the formation of the charge order is linked to the van Hove singularities  near the Fermi level~\cite{Liu2021, Kang2022,YHu2022}.  While it is suggested the van Hove singularities in CsV$_3$Sb$_5$ create a Fermi surface nesting scenario, in ScV$_6$Sn$_6$ we do not find any energy gaps at the Fermi level below T$_c$.  

On the other hand, the energy shift of the bands and Lifshitz transition does change the shape of the Fermi surface. The shift of spectral weight from the  $\bar{\text{M}}$ point towards the  $\bar{\text{K}}$ point across the CDW transition is similar to what is observed in the Fermi surface plots (Fig.~\ref{fig:FS}b, c).  However, we do not find any evidence that the CDW gaps the Fermi surface. Hence, the van Hove saddle point likely plays a role in the CDW formation, it is not a prototypical Peierls type instability. Nonetheless, the relocation of a large density of states at the Fermi level correlates with the reduced resistivity of the material as the CDW forms~\cite{Arachchige2022}. 

\begin{figure*}[t!]
\includegraphics[width=1.0\textwidth]{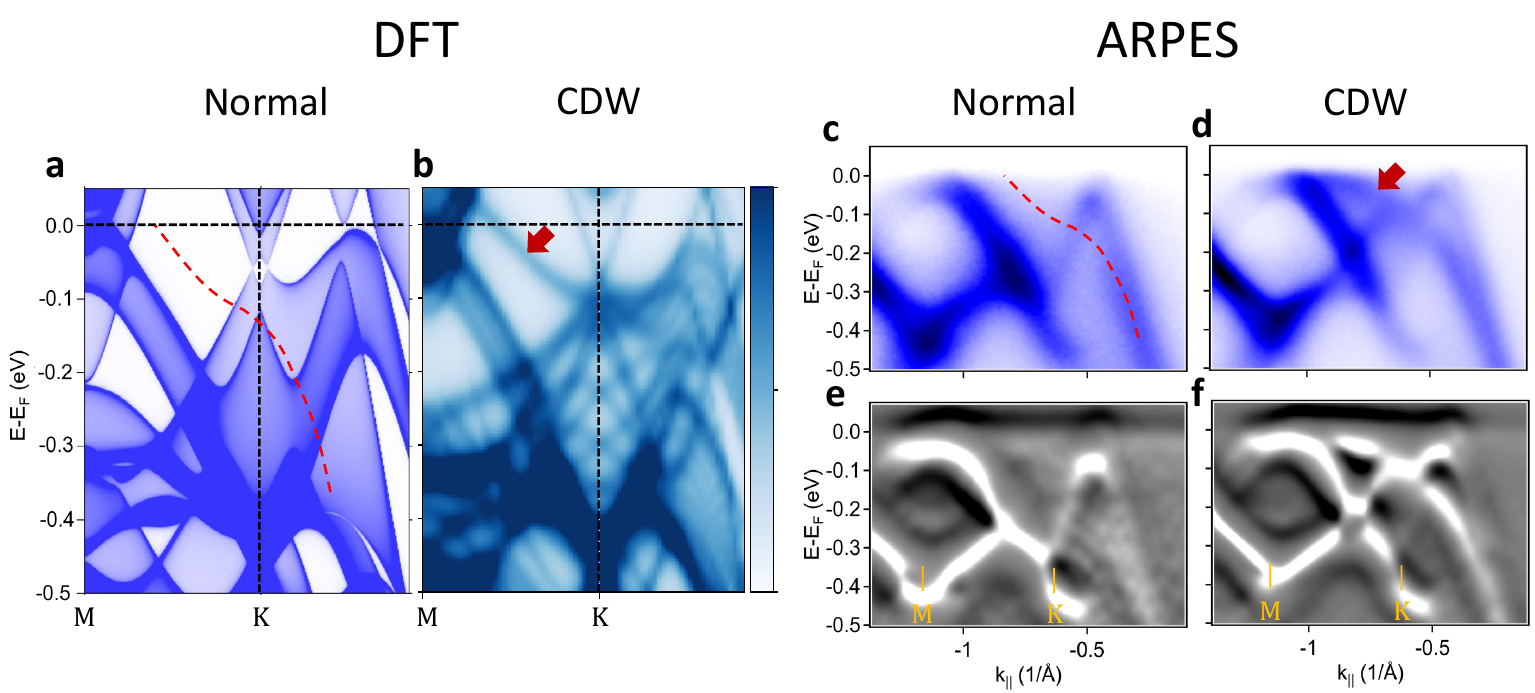}
\caption{Emergence of new bands from room temperature (normal state) to low temperature (CDW) structures.  a, DFT bands of the normal state and b, CDW state. c,d, ARPES data at T = 124~K in the normal state and at  T = 20~K in the CDW state, respectively. e,f, Curvature analysis for the normal and CDW states, respectively. The red dashed lines are to guide the eye to the absence and location of the extra band in the normal state that is present in the CDW state (red arrows). }
\label{fig:newband}
\end{figure*}

In our ARPES experiments the electronic structure in the normal and CDW phases are very similar, with one notable difference. In the low-temperature CDW phase an additional band near the Fermi level is observed that does not appear in the normal state as shown in Fig.~\ref{fig:newband}. Temperature dependent STM/S data also support this observation. %\textcolor{red}{and another study [https://arxiv.org/abs/2302.12227]}. 
Shown is Fig.~\ref{fig:SI-STM}c is the comparison of the local density of states (LDOS) below (4.6~K, blue) and above (120~K, black) the CDW transition temperatures from the kagome termination. There is a distinct peak around 50~mV below the Fermi level in the 4.6~K dI/dV map, but it is absent in the 120~K data. 

To find the origin of the additional band in the CDW phase, we first evaluated the band structures for low-temperature bulk structures. The band structures are investigated at $k_z=\frac{n}{6}$ ($n$=0, 1, 2, 3) due to the three times change in the structural periodicity along the c-axis by CDW. No additional bands were identified near the Fermi level with a similar band structure for high-temperature structures (normal state). (See Fig.~\ref{fig:SI-2}) 

We present the band structures of two kinds of surface terminations for the CDW state slab structures in the Supplementary Information (Fig.~\ref{fig:SI-slab}). To compare the electronic structures of the normal state between the calculation and ARPES experiments, we investigated unfolded band structures along the same high symmetry lines. For the Sn-termination (Fig.~\ref{fig:SI-slab}b,d), the dangling bond state of Sn on the surface appears above the Fermi level, but there is no additional band near the Fermi level. For the kagome termination (Fig.~\ref{fig:SI-slab}c,e), interestingly, we can clearly see an additional surface band analogous to the extra band observed in the ARPES experiment near the Fermi level. 

We also confirm that the orbital contribution of the additional band is the $d_{z^2}$ orbital of surface V atoms. We further analyze the evolution of additional bands by comparing the electronic structure for the kagome termination in the normal state and CDW state. In the band structures for the kagome termination in the normal state, the additional bands also exist with one unoccupied and one occupied band (see Fig.~\ref{fig:SI-1}g,h). As the temperature decreases below T$_c$, CDW is induced, and these additional bands move down below the Fermi level. It is a CDW transition from metal to metal, so it does not show the same energy gain as the band that went below the Fermi level as the band gap opened. However, it can be confirmed that the occupied band gains energy as it goes further down the energy level induced by CDW. 

Figures~\ref{fig:newband}c-f show Cut 1 with $h\nu$ = 98~eV at temperatures below and above the CDW transition, which emphasize the existence of the extra band in the CDW state. The extra band matches that found in the DFT calculation (Fig.~\ref{fig:newband}a,b). The differences are clearer in curvature method plots (Fig.~\ref{fig:newband}e,f) \cite{Zhang2011}. The extra band is missing in the normal state above T$_c$ as highlighted in Fig.~\ref{fig:newband}c. 
While this additional band in the CDW phase is the most prominent difference, there are more subtle changes in the electronic structure observed in the curvature method plots near the $\bar{\text{M}}$ point at $E$-$E_F$ = -0.4~eV and near the  $\bar{\text{K}}$ point at $E$-$E_F$ = -0.2~eV as highlighted in Fig.~\ref{fig:SI-nogap}. Additional data for the evolution of the electronic structure at intermediate temperatures is provided in the Supplementary Information Figs.~\ref{fig:SI-3},~\ref{fig:SI-4},~and ~\ref{fig:SI-5}.

\section{Conclusion}
Our study of the topological kagome metal, ScV$_6$Sn$_6$, has revealed a novel phenomenon of surface bands and Lifshitz transition which are evident at the critical temperature of the CDW state. This finding is significant as it provides new insights into the electronic properties of kagome materials and their CDW behavior. The identification of this new surface band could also lead to the design and development of new materials with unique functionality. Future studies on kagome materials may benefit from the findings of our investigation, which sheds light on the origin of electronic instabilities in these unique systems.

\section*{Methods}
Our ab \textit{initio} calculations are based on density functional theory (DFT) \cite{HK1964,KS1965} as implemented in the Vienna ab \textit{initio} simulation package (VASP) \cite{Kresse1993,Kresse1996} with projector augmented wave potentials \cite{PAW1994,Kresse1999} and spin-orbit coupling. The Perdew-Burke-Ernzerhof (PBE) form is employed for the exchange-correlation functional with the generalized gradient approximation (GGA) \cite{GGA1996}. The energy cutoff is set to 520~eV for all calculations.
For the crystal structures, we used the P6/mmm space group with lattice constants, a,b = 5.4530~\AA and c=9.2311~\AA, and 13 atoms for the high-temperature (normal state) structure. We use the experimental structure \cite{Arachchige2022} for P6/mmm space group supercell ($\sqrt{3}\times\sqrt{3}\times$3) in the low-temperature structure with lattice constants, $a=b=9.4433$~\AA~and $c=27.7281$~\AA, and 117 atoms. The Brillouin zone is sampled using a 31$\times$31$\times$11 $\Gamma$-centered $k$-grid for the high-temperature structure, and 11$\times$11$\times$3 $\Gamma$-centered $k$-grid for the low-temperature structure, respectively. From the DFT result, we obtain the maximally localized Wannier functions for 4$s$ and 3$d$-orbitals of V atom and 5$p$-orbitals on Te atoms by using the WANNIER90 code \cite{Mostofi2014}, which are used to analyze the surface density of states. The surface projected local density of states was calculated by the WANNIERTOOLS \cite{Wu2018}, which is based on the iterative Green’s function technique \cite{Sancho1985}.

ScV$_6$Sn$_6$ crystals were grown from a Sn-rich melt as described in our recent paper~\cite{Arachchige2022}. The crystals grow as hexagonal blocks 0.4-3mm in size. ScV$_6$Sn$_6$ exhibit relatively poor 001 cleavage in contrast to the excellent 001 cleavage we observe in LuV$_6$Sn$_6$, YV$_6$Sn$_6$ or RMn$_6$Sn$_6$ crystals.

Synchrotron based ARPES measurements were carried out at Beamline 4.0.3 at the Advanced Light Source utilizing a Scienta R8000 photoelectron analyzer allowing for an angular resolution less than 0.2$^{\circ}$ and an energy resolution better than 20~meV.  Samples were cleaved in vacuum at a base pressure better than 5$\times$ 10$^{-11}$ Torr.  Measurements were performed with both linearly horizontal and linearly vertical polarizations in the energy range $h\nu$ = 38 – 124~eV and a sample temperature range of T = 20 – 124~K.

Single crystals were cleaved in ultrahigh vacuum (UHV) at low temperature and then immediately transferred to the scanning tunneling microscopy/spectroscopy (STM/S) head which was precooled to 4.6~K or 78~K without breaking the vacuum. The STM/S experiments were carried out using a UHV STM with a base pressure better than 2 $\times$ 10$^{-10}$~Torr. Pt-Ir tips (electro-polished after mechanical grinding) were conditioned on Au(111) surface before each measurement.

\section*{Acknowledgement} 
Theory work and ORNL-led synchrotron based ARPES measurements were supported by the US Department of Energy, Office of Science, Office of Basic Energy Sciences, Materials Sciences and Engineering Division (J. W. V., M. Y., H. L. and R. M.), and by the U.S. Department of Energy (DOE), Office of Science, National Quantum Information Science Research Centers, Quantum Science Center (S.-H. K. and Q. L.). 
This research used resources of the Advanced Light Source, which is a DOE Office of Science User Facility under contract No. DE-AC02-05CH11231. D.M. acknowledges support from the US Department of Energy, Office of Science, Basic Energy Sciences, Materials Science and Engineering Division. H.W.S.A and W.R.M. acknowledge support from the Gordon and Betty Moore Foundation’s EPiQS Initiative, Grant GBMF9069 to DM. STM/S research conducted at the Center for Nanophase Materials Sciences (CNMS), which is a US Department of Energy, Office of Science User Facility at Oak Ridge National Laboratory (S. H., H. J., and Z. G.). This research used resources of the Oak Ridge Leadership Computing Facility and the National Energy Research Scientific Computing Center, US Department of Energy Office of Science User Facilities.

This manuscript has been authored by UT-Battelle, LLC, under Contract No. DE-AC0500OR22725 with the U.S. Department of Energy. The United States Government retains and the publisher, by accepting the article for publication, acknowledges that the United States Government retains a non-exclusive, paid-up, irrevocable, world-wide license to publish or reproduce the published form of this manuscript, or allow others to do so, for the United States Government purposes. The Department of Energy will provide public access to these results of federally sponsored research in accordance with the DOE Public Access Plan (http://energy.gov/downloads/doe-public-access-plan).

\clearpage
\bibliographystyle{apsrev4-2}
\bibliography{biblio}
\clearpage

\renewcommand{\thefigure}{S\arabic{figure}}
\setcounter{figure}{0}

\section*{Supplemental information} 
%\counterwithin{figure}{section}

\subsection{Scanning tunneling microscopy and spectroscopy}
\begin{figure*}[htb!]
\includegraphics[width=1.0\textwidth]{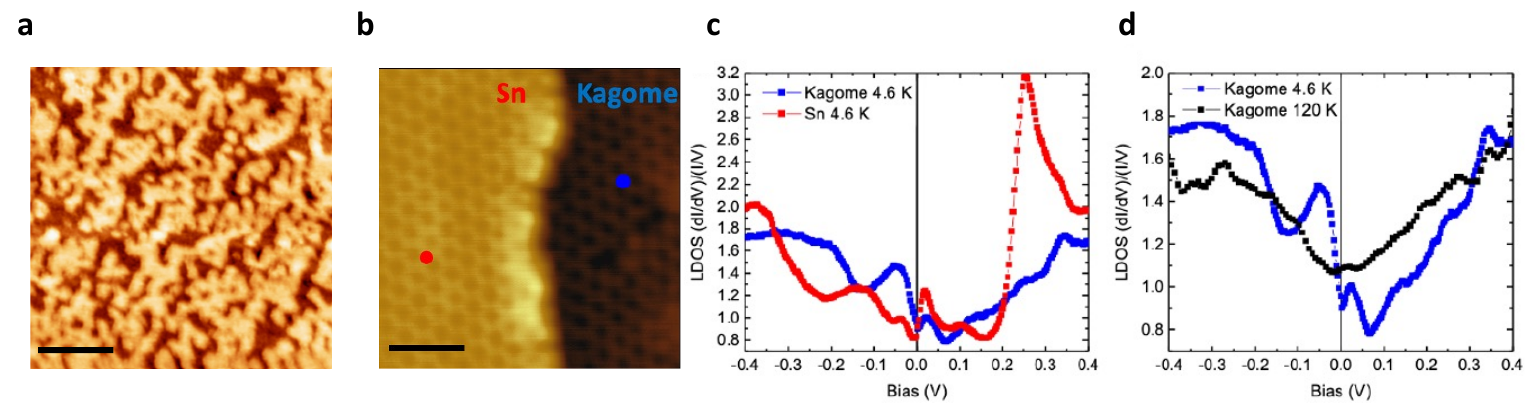}
\caption{STM/S of cleaved ScV$_6$Sn$_6$ Surface. a, STM image of ScV$_6$Sn$_6$ surface cleaved at room temperature shows kagome terminated terrace (dark regions) together with Sn terminated island (bright regions) with scale bar 20~nm. b, STM image of a typical area of ScV$_6$Sn$_6$ cleaved at low temperature, shows the coexisting kagome (right side) and Sn (left side) terraces. The measured step height is 0.18 $\pm$ 0.04~nm. The scale bar is 2~nm. The blue and red dots show approximate locations for the blue and red spectra in panel c. c, distinct LDOS spectra for the kagome (blue) and Sn (red) terminations at 4.6~K. d, Comparison of the LDOS below (4.6~K, blue) and above (120~K, black) T$_c$ on the kagome termination. The set point for panel a is 35~mV, 0.5~nA, b is -0.2~V, 0.5~nA, panel c is -0.2~V, 0.5~nA, and panel d is -0.75~V, 0.5~nA.}
\label{fig:SI-STM}
\end{figure*}

Scanning tunneling microscopy and spectroscopy (STM/S) of the cleaved crystals at room temperature and low temperature display a mixed surface with coexisting kagome and Sn terminations.  Fig.~\ref{fig:SI-STM}a,b show a typical STM image of the in-situ cleaved ScV$_6$Sn$_6$ surface, which shows the coexisting kagome (dark region) and Sn (bright region terraces, as identified by the step height of 0.18 ± 0.04~nm. The difference of the two terminations is also proved by the distinct local density of the states [calculated from (dI/dV)/(I/V), i.e., normalized dI/dV] of the two surfaces as shown in Fig.~\ref{fig:SI-STM}c, for the kagome (blue) and Sn (red) terminations at 4.6~K. 

\subsection{Analysis of energy distribution curves (EDCs) along high symmetry points}
\begin{figure*}[htb!]
\includegraphics[width=1.0\textwidth]{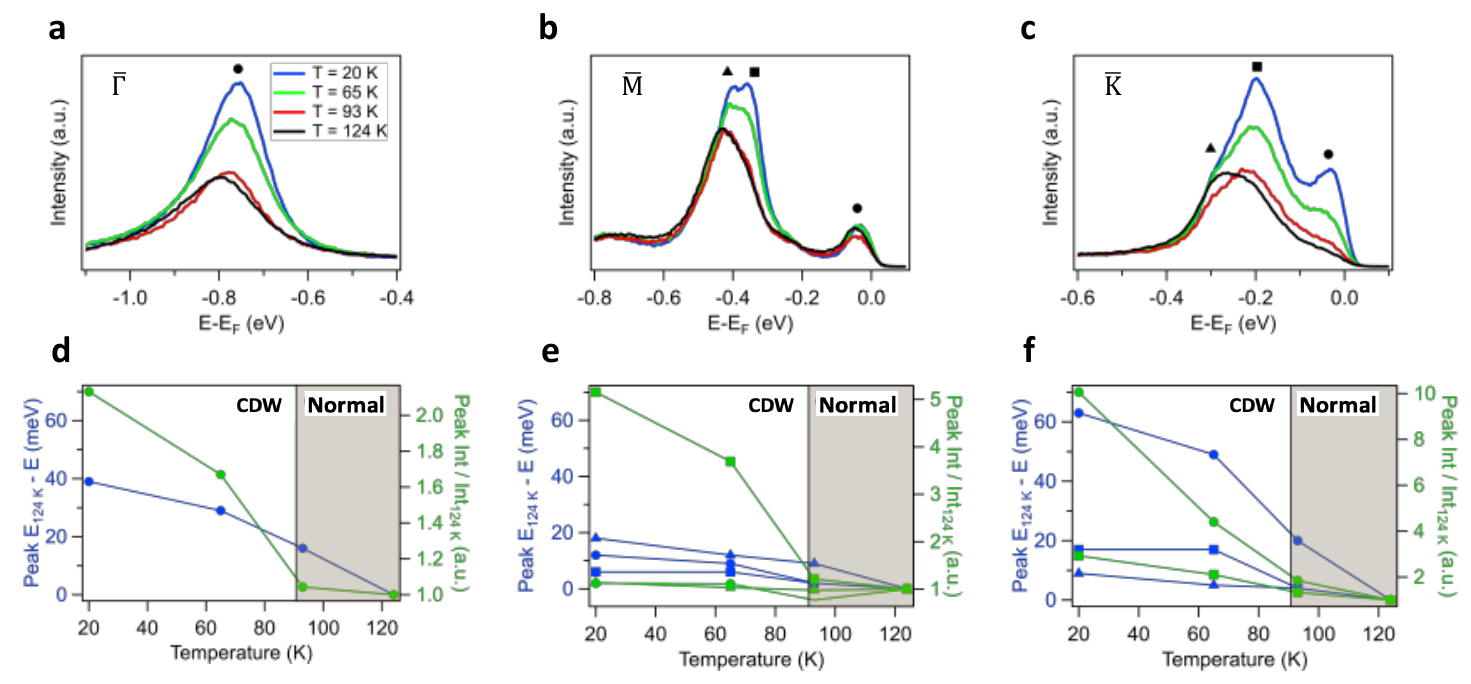}
\caption{Electronic structure analysis along high symmetry cuts.  Energy distribution curves (EDC) at different temperatures at a-c, $\bar{\Gamma}$, $\bar{\text{M}}$, and $\bar{\text{K}}$, respectively. Markers highlight peaks used in Lorentz fits of the data. d, Lorentz fit of peak position and relative intensity for the one band highlighted at $\bar{\Gamma}$.  e, Lorentz fits of peak positions and relative intensities of the three bands highlighted at $\bar{\text{M}}$. f, Lorentz fits of peak positions and relative intensities of the three bands highlighted at $\bar{\text{K}}$.}
\label{fig:SI-nogap}
\end{figure*}

To better understand the additional band as well as the more subtle changes to the electronic structure across the CDW phase transition, energy distribution curves (EDCs) through high symmetry points are provided in Fig.~\ref{fig:SI-nogap}.  The data are normalized to a background region away from any dispersive feature for comparison of relative changes to the features and then fit with Lorentz functions to track band position and intensity with temperature.  In general, there appears a $\sim$40 meV upward shift of the entire electronic structure as the sample is warmed from T = 20 K to T = 124 K, as represented by the EDC cut through the $\bar{\Gamma}$ point.  While similar trends are observed for the band structure in general, the EDCs cuts through the $\bar{\text{M}}$ and $\bar{\text{K}}$ points reveal changes in the line shapes with temperature indicative of changes in the band dispersions or band character at these points.  As highlighted in the Fig.~\ref{fig:SI-nogap}b,e at the $\bar{\text{M}}$ point, the band positions for the two features at $E$-$E_F$ = -0.4 eV change relative to each other, suggestive of a small $\sim$10 meV gap opening as the temperature decreases across T$_c$.  In addition, the band marked with the square shows a significant increase in intensity while the band marked with the triangle shows a near constant intensity as the system enters the CDW phase.  The feature at the Fermi level shows a smaller shift with temperature compared to the $\bar{\Gamma}$ band.  However, the magnitude of this change could be skewed due to the band being truncated at the Fermi level.  Similar behaviors are observed for the two band features at $E$-$E_F$ = -0.2 eV in the $\bar{\text{K}}$ EDC shown in Fig.~\ref{fig:SI-nogap}c,f, i.e. a small $\sim$8-10 meV gap appears to open while the feature marked with a square has a significant increase in intensity relative to the other band as the CDW phase is entered. The most prominent change in the $\bar{\text{K}}$ EDC is the $\sim$60 meV shift in band position and factor 10 increase in band intensity for the band near $E_F$.  This particular band in the $\bar{\text{K}}$ EDC illustrates the appearance of the additional band noted in Fig.~\ref{fig:newband} as the temperature is cooled below the CDW phase transition.

\begin{figure*}[htb!]
\includegraphics[width=0.6\textwidth]{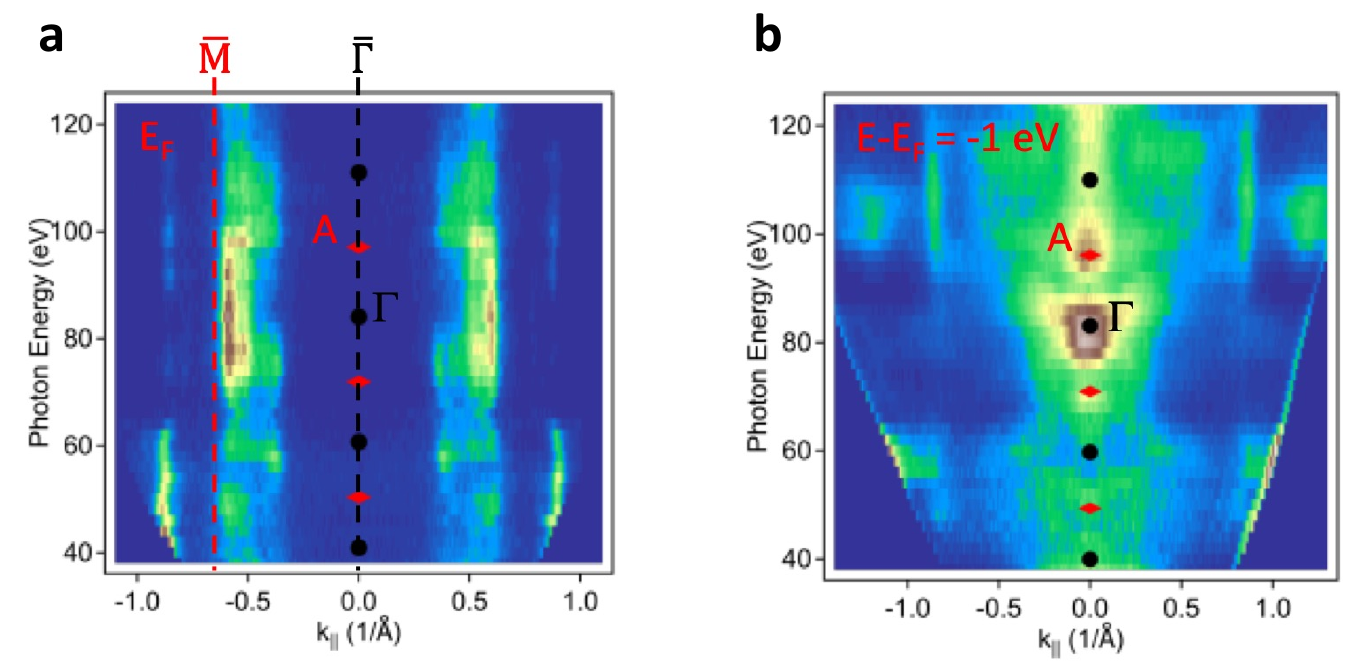}
\caption{Analysis of $k_z$ dispersion. (a) Measurement of $T = 20$ K electronic structure centered along $\bar{\Gamma}-\bar{\text{M}}$ at $E$-$E_F$ = -1.0 eV for different photon energies. Brillouin zone high symmetry points determined using $c = 9.2$~\AA~with an empirical inner potential of 8.5 eV and estimated work function of 4.5 eV. (b) Measurement of electronic structure along same cut as (a) but at $E_F$.  The data show little warping of the bands due to $k_z$ dispersion but do show intensity modulations due to photon dependent matrix element effects.}
\label{fig:SI-kz}
\end{figure*}

\begin{figure*}[htb!]
\includegraphics[width=1.0\textwidth]{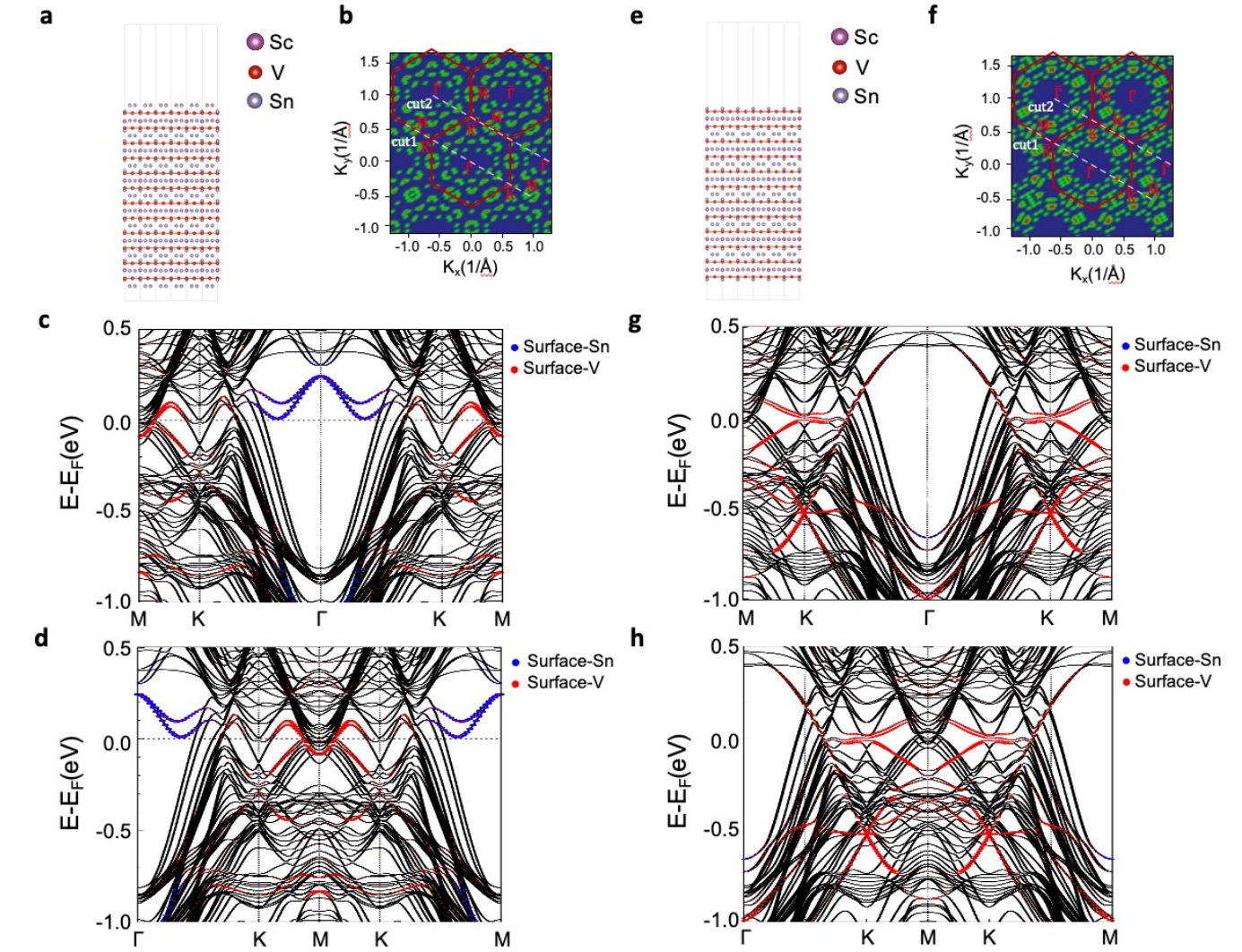}
\caption{Electronic structures of high-temperature slab structure in ScV$_6$Sn$_6$. a, Sn-terminated atomic structure. b, Calculated Fermi surface of Sn-terminated slab. c,d,  Calculated band structures for the Sn-terminated slab along high symmetry lines. e, kagome-terminated atomic structure. f, Calculated Fermi surface of kagome-terminated slab. g,h, Calculated band structures for the kagome-terminated slab along high symmetry lines.}
\label{fig:SI-1}
\end{figure*}

\begin{figure*}[htb!]
\includegraphics[width=0.7\textwidth]{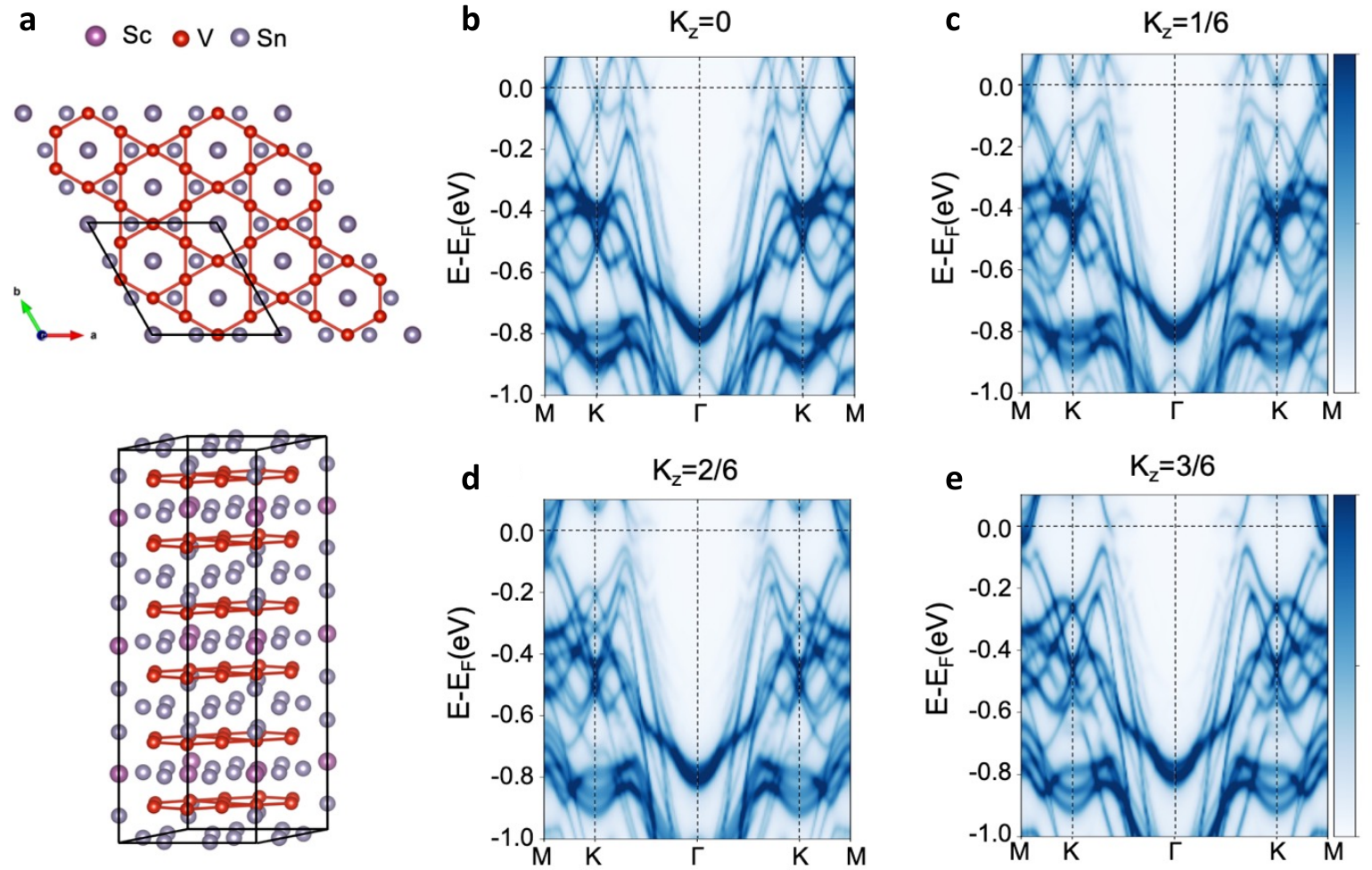}
\caption{Electronic structures of low-temperature bulk structure (CDW phase) in ScV$_6$Sn$_6$. a, Low-temperature bulk atomic structure, b-e, Calculated unfolded band structures for the low-temperature bulk along high symmetry line at $k_z=0$, 1/6, 2/6, and 3/6, respectively.}
\label{fig:SI-2}
\end{figure*}

\begin{figure*}[htb!]
\includegraphics[width=0.7\textwidth]{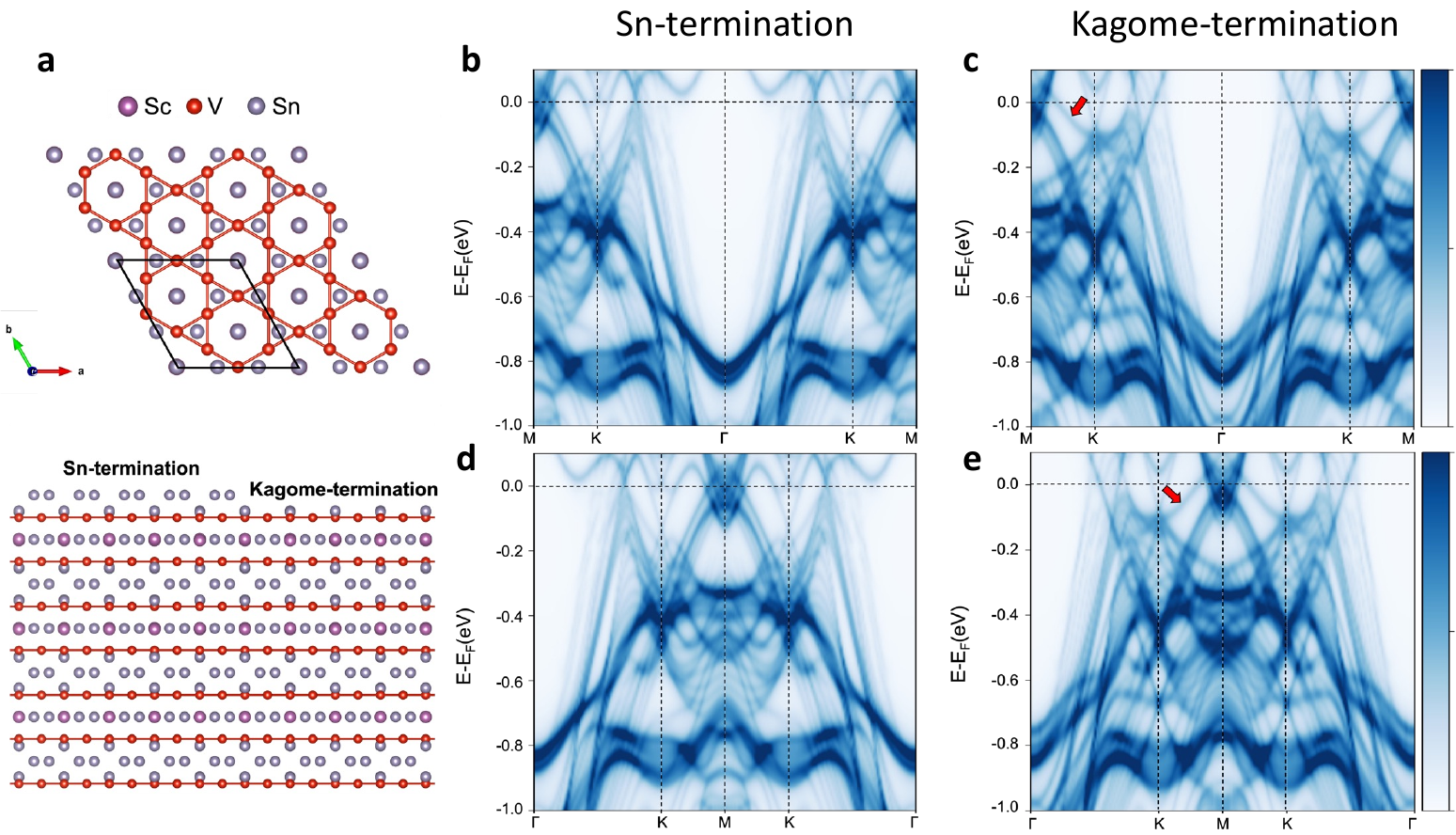}
\caption{Electronic structures of low-temperature slab structures (CDW phase) in ScV$_6$Sn$_6$. a, Low-temperature atomic slab structures for Sn-termination and kagome-termination, b,d, Calculated unfolded band structures for the low-temperature slab for Sn-termination and, c,e, kagome-termination along high symmetry lines, respectively. The extra kagome-surface state is indicated by the red arrow.}
\label{fig:SI-slab}
\end{figure*}

\begin{figure*}[htb!]
\includegraphics[width=1.0\textwidth]{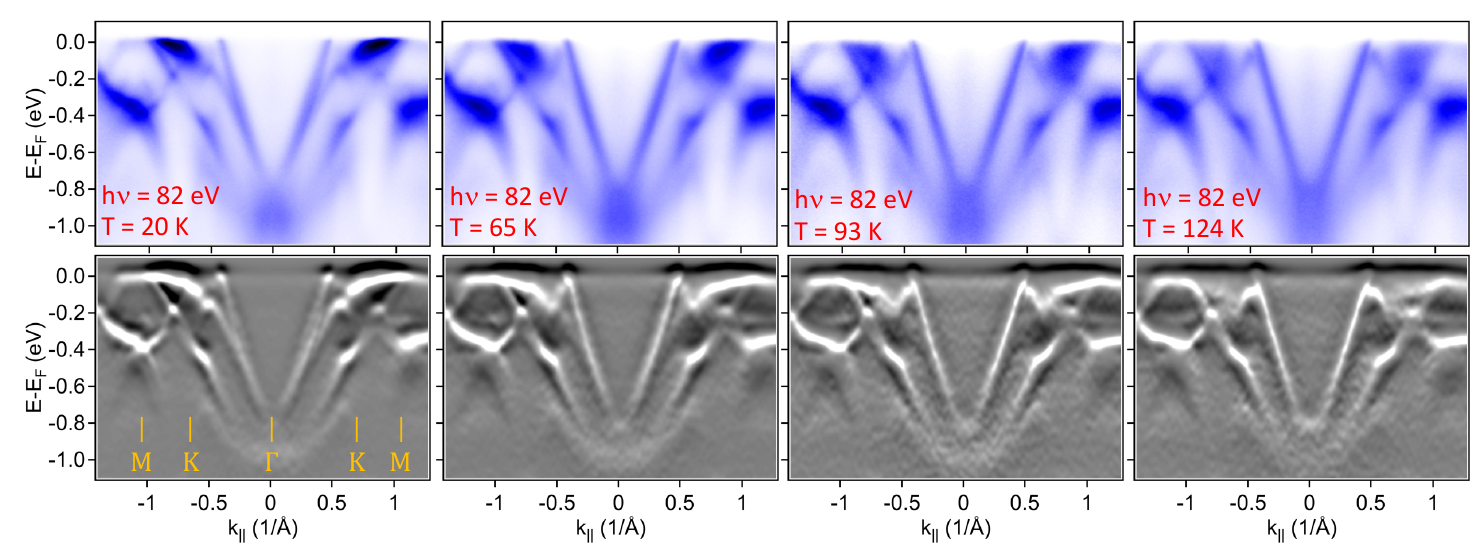}
\caption{Temperature dependent data with photon energy $h\nu$ = 82 eV.  ARPES data (top) and curvature analysis (bottom) for data taken at different temperatures as annotated in the ARPES panels.  High symmetry points for all the panels highlighted in the left bottom panel.}
\label{fig:SI-3}
\end{figure*}

\begin{figure*}[htb!]
\includegraphics[width=1.0\textwidth]{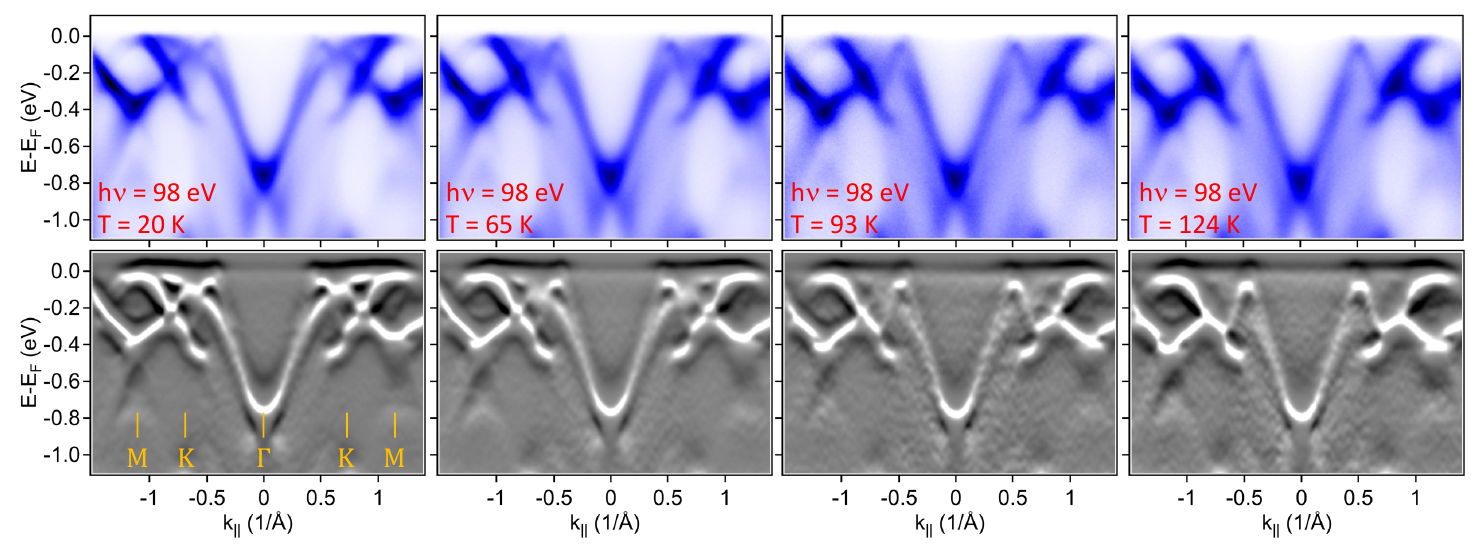}
\caption{Temperature dependent data with photon energy $h\nu$ = 98 eV.  ARPES data (top) and curvature analysis (bottom) for data taken at different temperatures as annotated in the ARPES panels.  High symmetry points for all the panels highlighted in the left bottom panel.}
\label{fig:SI-4}
\end{figure*}

\begin{figure*}[htb!]
\includegraphics[width=1.0\textwidth]{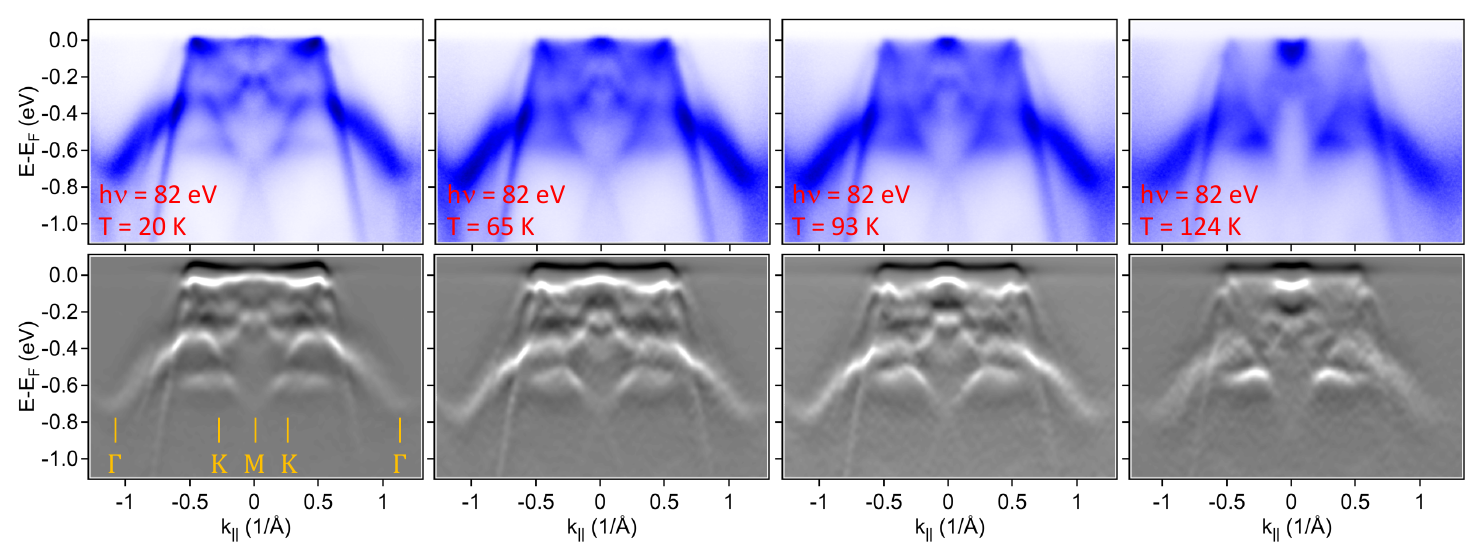}
\caption{Temperature dependent data with photon energy $h\nu$ = 82 eV.  ARPES data (top) and curvature analysis (bottom) for data taken at different temperatures as annotated in the ARPES panels.  High symmetry points for all the panels highlighted in the left bottom panel.}
\label{fig:SI-5}
\end{figure*}
\end{document}